%% file: finetuning.tex
\documentclass[nopreprintnumbers,apsd,twocolumn,twoside]{revtex4}
\usepackage{graphicx,amsmath,amssymb,latexsym}

 \newcommand{\f}[2]{\frac{#1}{#2}}

\newcommand{\be}{\begin{equation}} \newcommand{\ee}{\end{equation}}
\newcommand{\barr}{\left(\begin{array}}
\newcommand{\earr}{\end{array}\right)}

\begin{document}
\pagestyle{myheadings} \markboth{Persistent Fine-Tuning in
Supersymmetry and the NMSSM}{Persistent Fine-Tuning in Supersymmetry
and the NMSSM}

\title{{\bf Persistent Fine-Tuning in Supersymmetry and the NMSSM}}
\author{Philip C. Schuster\footnote{email: schuster@fas.harvard.edu} and Natalia Toro\footnote{email: toro@fas.harvard.edu}}
\affiliation{Jefferson Laboratory of Physics, Harvard University,
Cambridge, Massachusetts 02138, USA} \preprint{unknown}
\begin{abstract}
\input{abstract.tex}
\end{abstract}
\maketitle


\input{intro.tex}
\input{TuningOverview.tex}
\input{NmssmOverview.tex}
\input{conclusion.tex}

\section{Acknowledgements}
P.C.S. and N.T. would especially like to thank Nima Arkani-Hamed for
inspiring this project and for many insightful discussions and
comments. Additional thanks to Andre Sopczak for providing us with
recent combined LEP $H\rightarrow 4\mbox{b}$ exclusion data.
Additional thanks to Spencer Chang, Aaron Pierce, Jesse Thaler, and
Lian Tao Wang for many helpful discussions. N.T. and P.C.S. are each
supported by NDSEG Fellowships.

\appendix

\end{document}

%% file: abstract.tex
\noindent We examine the use of modified Higgs sectors to address
the little hierarchy problem of supersymmetry.  Such models reduce
weak-scale fine-tuning by allowing lighter stops, but they retain
some fine-tuning independent of the Higgs.  The modified Higgs
sector can also introduce model-dependent tunings. We consider the
model-independent constraints on naturalness from squark, gluino,
chargino and neutralino searches, as well as the model-dependent
tuning in the PQ- and R-symmetric limits of the NMSSM, where cascade
decays or additional quartic interactions can hide a Higgs without
relying on heavy stops. Obtaining viable electroweak symmetry
breaking requires a tuning of $\sim 1-10\%$ in the PQ limit. A large
A-term is also necessary to make charginos sufficiently heavy, and
this introduces an additional weak-scale tuning of $\sim 10\%$. The
R-symmetric limit requires large marginal couplings, a large singlet
soft mass, and a tuning of $\sim 5-10\%$ to break electroweak
symmetry. Hiding MSSM-like Higgs states below $\approx 112$ GeV
requires additional tunings of A-terms at the $\sim 1-10\%$ level.
Thus, although the NMSSM has rich discovery potential, it suffers
from a unique fine-tuning problem of its own.

%% file: intro.tex
\section{Introduction}\label{sec:intro}
The little hierarchy problem of supersymmetry is a blight on an
otherwise impressive new-physics framework. Supersymmetry may
explain both the mechanism of electroweak symmetry breaking and the
weakness of gravity \cite{SUSY Reference}. Moreover, in the minimal
supersymmetric Standard Model (MSSM), gauge coupling unification
occurs with striking accuracy near $M_{GUT}\approx 2\times 10^{16}$
GeV and the lightest superpartner is a cold dark matter candidate.

Yet the heaviness of superpartners has created an apparent
fine-tuning problem. $M_Z^2$ is determined by the Higgs soft masses,
and depends indirectly on stop and gluino masses. A hierarchy
between the $Z$ and the stop and gluino masses implies fine-tuning
of the spectrum. The non-discovery of a Higgs at LEP, which is only
consistent with the MSSM if the stop mass is $\gtrsim 500$ GeV,
hints at such a hierarchy. For this reason, many attempts to reduce
this little hierarchy have focused on the Higgs sector, either hiding
a light Higgs or adding new Higgs quartic interactions so that
$m_H>M_Z$ is consistent at tree-level.

But two sources of fine-tuning still plague models with a hidden or
heavy Higgs.  First, the Higgs is just one of many indications of a
$Z$-stop hierarchy and the resulting tuning---the absence of other
superpartners at LEP and Tevatron also implies a small hierarchy.
Second, the modified Higgs sectors may also require additional
fine-tuning to be consistent with data. The latter question is
model-dependent, and we consider it in the context of the NMSSM.

In Section \ref{sec:tuning} we discuss the implications of
experimental limits on superpartner masses for fine-tuning of the
$Z$-mass. These constraints come not only from Higgs exclusions, but
also from gluino, squark, chargino and neutralino bounds. Abandoning
standard theoretical assumptions can reduce the fine-tuning implied
by these results, but at the cost of greater complexity and lessened
predictive power.

In Section \ref{sec:nmssm}, we consider the impact of a minimally
extended Higgs sector (the NMSSM) on fine-tuning.  In the context of
the NMSSM, lifting the Higgs mass through new quartic couplings and
opening up cascade decay channels for a light Higgs so that decay to
$b\bar b$ is suppressed \cite{Dermisek:2005ar, Other NMSSM hidden
higgs references} have been proposed as alternate means of
explaining the non-discovery of the Higgs, alleviating fine-tuning.
Either scenario can potentially be natural in the limit of an
approximate $U(1)_{PQ}$ or $U(1)_R$ symmetry.  Away from these
limits, electroweak symmetry breaking is tuned and the Higgs sector
must be tuned to allow cascade decays.

The pseudo-Goldstone boson of the broken $U(1)_{PQ}$ naturally opens
a cascade decay channel that can hide the lightest MSSM-like Higgs
below $\approx 110$ GeV for $\kappa \lesssim 0.05$. Large $\lambda
\gtrsim 0.65$ and $A_\lambda \gtrsim 300-350$ are required to raise
the singlino and higgsinos above LEP bounds. The large $A$-term
reintroduces a weak scale tuning of $\sim 10\%$. Electroweak
symmetry only breaks when the up-type and singlet soft masses are
made unnaturally small, thereby introducing an additional tuning of
$\sim 1-10\%$.

In the $U(1)_R$ limit, severe tuning of the spectrum and mixing
angles is required to hide an MSSM-like Higgs below $\approx 112$
GeV. The A-terms must be much smaller than typical radiative
corrections, and so a tuning of $\sim 1-10\%$ is required. This is
not true of the points discussed in \cite{Other NMSSM hidden higgs
references}, which naturally explain the excess of $2b$ events
associated with a Higgs mass of $\approx 100$ GeV
\cite{LEPmssmHiggs}. Rather, the tuning we discuss is what is
required for the Higgs to go undetected, generically, given the
present experimental constraints.  For large singlet sector
couplings where the Higgs states are heavier, direct search
constraints are more readily avoided, but a tuning of $\sim 10\%$ is
still required for electroweak symmetry to break.

Therefore, the NMSSM suffers from its own fine-tuning problem. The
discrepancy between these conclusions and the finding in
\cite{Dermisek:2005ar} of points with mildly tuned $M_Z$ arises from
their narrower notion of fine-tuning and broader focus in parameter
space. Points considered in \cite{Dermisek:2005ar} outside of the PQ
or R limits may indeed have less fine-tuned $Z$ masses, but they
ignore the tunings required for viable electroweak symmetry breaking
and disregard tunings of Higgs branching ratios and pseudoscalar
masses necessary to hide the Higgs.

Because the strongest constraints on the NMSSM PQ limit are related
not to the Higgs bosons but to electroweak symmetry breaking and
higgsino masses, the fine-tuning may be reduced by adding additional
couplings.  This could present a new avenue for high-scale SUSY
model-building, parallel to that of explaining the relative lightness
of the $Z$ compared to superpartners.

%% file: TuningOverview.tex
\section{Sources of Fine-Tuning in SUSY}\label{sec:tuning}
In supersymmetric extensions of the Standard Model, technical
naturalness demands several relations between mass scales. In
particular,
\begin{eqnarray}
\label{eq:natural1}  m_Z^2 &\sim& m_{H_u}^2 ,\\
\label{eq:natural2}  m_{H_u}^2 &\sim& m_{\tilde t}^2 \sim m_{\tilde g}^2 ,\\
\label{eq:natural3}  (m_{\tilde g} &\sim& 3-6 m_{\tilde \chi}),
\end{eqnarray}
where $m_{H_u}, m_{H_d}$ are the Higgs soft masses, and $m_{\tilde
t}$, $m_{\tilde g}$, and $m_{\tilde \chi}$ are the physical mass
scales for the stops, gluinos, and neutralinos/charginos
\cite{Martin:1997ns}.

These relations suggest that the $Z$ should be found near the middle
or even the top of the superpartner spectrum---an expectation that
has not been borne out by experiment. Fine-tuning is required to
make the $Z$ light compared to superpartners, and is minimized when
the superpartners are as light as possible or when the relations
(\ref{eq:natural1})--(\ref{eq:natural3}) do not apply.

Relation (\ref{eq:natural1}) above follows from
minimization of the Higgs potential, from which
\begin{equation}\label{HiggsMinimization}
m_Z^2 = \f{1-\cos(2\beta)}{\cos(2\beta)}m_{H_u}^2-
  \f{1+\cos(2\beta)}{\cos(2\beta)}m_{H_d}^2-2\mu^2
\end{equation}
where
$\tan(\beta)=\f{v_u}{v_d}$. Relation (\ref{eq:natural2}) stems from
RG-dependence of $m_{H_u}^2$ on the soft masses $m_{Q3}^2, m_{U3}^2$
controlling $m_{\tilde t}^2$ and of $m_{Q3}^2, m_{U3}^2$ on
$m_{\tilde g}^2$ which are enhanced by the large couplings $y_t$ and
$g_3$. The one-loop corrections to these soft masses are
\begin{eqnarray}
  \delta m_{H_u}^2   &\approx& -\f{3y_t^2(m_{Q3}^2+m_{U3}^2)}{8\pi^2}\ln(\f{\Lambda}{m_S}), \nonumber \\
\label{RadiativeCorrA}  \delta m_{Q3,U3}^2 &\approx&
  \f{8\alpha_3M_3^2}{3\pi}\ln(\f{\Lambda}{m_S}),
\end{eqnarray}
where $M_3$ is the gluino soft mass, and $\Lambda$ and $\tilde m_S$
are the mediation scale and superpartner mass scale respectively
\cite{Martin:1997ns}. If the cutoff of the theory is high (e.g.
$\Lambda\sim M_{GUT}$), then the logarithmic enhancement nearly
cancels the loop factors. Relation (\ref{eq:natural3}) follows from
the assumption of universal gaugino masses, with the numerical
factors derived from ratios of gauge couplings.

Fine-tuning in SUSY is often attributed to the non-discovery of a
light Higgs at LEP, but in fact this is only one of several sources
of tension. The non-observation of gluinos and squarks below $\sim
200-300$ GeV suggests tuning at the $10-20\%$ level or worse.
If gaugino masses unify, the non-observation of neutralinos and
charginos leads to even greater tuning.

Additional constraints on SUSY also appear as fine-tunings.  We do not
consider here the tuning required for reasonable dark matter abundance
(below closure density) or the smallness of non-minimal flavor
violation.  These are largely decoupled from the tuning problems we
consider here.

Two sources of tension that are highly dependent on the Higgs sector,
which we will discuss in the next section, are stable electroweak
symmetry breaking and Higgs and higgsino phenomenology with modified Higgs
sectors.

This category of tuning has received little attention in the
literature, but is morally similar to the familiar $Z$-mass
fine-tuning.  Such fine-tuning occurs when phenomenology requires a
cancellation among larger and unrelated parameters. We can
parameterize $Z$-mass tuning involving a parameter $a$ by $T\sim
(\f{d\log{M_Z^2}}{d\log{a}})^{-1}$. More generally, if a criterion
$f(a,b)<f_c$ imposes a constraint on $a$ and $b$, then the local
variation allowed without exceeding $f_c$ defines fine-tuning $T\sim
d\log{a}$, which quantifies the degree of cancellation between
parameters necessary to satisfy a criterion.

\subsection{Tension with a Heavy Higgs}
At tree-level, the lightest MSSM Higgs mass is bounded by $m_H<M_Z
|\cos 2\beta|$, yet masses $m_H<114$ GeV are excluded by LEP
\cite{LEPmssmHiggs}. Lifting $m_H$ above $M_Z$ at one-loop requires
a large stop loop contribution to $m_H$ given by
\cite{Martin:1997ns}
\begin{equation}\label{HiggsMassCorrection}
    \delta m_H^2=\frac{3}{4 \pi^2}v^2 y_t^4
    \sin^4 \beta \ln\left(\frac{m_{\tilde t_1} m_{\tilde t_2}}{m_t^2}\right),
\end{equation}
which pushes the Higgs above the LEP bound for $m_{\tilde t}\gtrsim
500$ GeV. If SUSY breaking is mediated at a moderately high scale,
then by eq. (\ref{RadiativeCorrA}) $m_{H_u}^2$ also receives
radiative corrections of $\sim -(400)^2$ GeV$^2$ that must be
cancelled by a tree-level term. This implies a tuning worse than
$5\%$.

This tuning can be remedied in several ways, of which we list only a
few. (i) The Higgs mass can be protected from tree-level sources, so
that it is naturally lighter than sparticles by a loop factor and
the tree-level relation (\ref{HiggsMinimization}) is irrelevant
\cite{One Loop Separation}. (ii) The Higgs mass can be increased by
new contributions to its quartic coupling beyond the SM D-term
source \cite{Adding Quartics}. (iii) The Higgs decay modes can be
modified so that LEP and CDF are insensitive to a Higgs below $114$
GeV \cite{Chang:2005ht}. (iv) The SUSY-breaking mediation scale can
be lowered, making $m_{H_u}^2$ less sensitive to $m_{Q3}$ and
$m_{U3}$ \cite{Low SUSY Scale}. Other approaches are developed in
\cite{Other LHP Solutions}.

All of these cures, especially (ii) and (iii), can require
additional tuning to produce viable phenomenology or stable
electroweak symmetry breaking.  As we discuss in the next section,
this must be counted against the gain in naturalness of $M_Z$.
Options (i) and (iv) often require significant and seemingly \emph{ad
  hoc} extensions of the MSSM that may introduce new hidden
tunings.

\subsection{Tension with Heavy Gluinos and Squarks}
Direct gluino and first- and second-generation squark searches
constrain the soft masses $m_{Q3}$, $m_{U3}$, and $m_{\tilde g}$
regardless of the Higgs mass. If the gluino is heavy ($m_{\tilde
g}\gtrsim 500$ GeV), squark masses as low as $m_{\tilde q}\gtrsim
200$ GeV are experimentally allowed. For heavier squarks, $m_{\tilde
q} \gtrsim 380$ GeV, the gluino can be as light as $m_{\tilde g}
\gtrsim 230$ GeV \cite{Gluino Squark}. If $m_{Q3,U3} \sim 330$ GeV
as well, then the $Z$ mass is tuned at the $10\%$ level or worse.

Constraints on the soft masses for the third-generation squarks are
much weaker than for the first two generations.  This is
particularly true for light-gluino scenarios, because only
first-generation squarks can be produced through $t$-channel
processes mediated by a gluino. Direct sbottom searches imply
$m_{\tilde b}\gtrsim 200$ if the sbottom decays through a neutralino
with $m_{\tilde \chi_1^0}\lesssim 80$ GeV \cite{Sbottoms}. If the
gluino is light, $m_{\tilde g}\lesssim 280$ GeV, then the constraint
becomes a little stronger ,$m_{\tilde b}\gtrsim 230$ GeV
\cite{Sbottoms}. Conservatively, this implies $m_{Q3}^2\gtrsim
(230)^2$ GeV$^2$.

The bound $m_{\tilde t}\gtrsim 150$ GeV from direct stop searches
does not constrain the soft masses at all, as it is saturated by the
contribution from $m_t$ \cite{StopD0}. However, we expect
$m_{U3,Q3}$ to be close to $m_{\tilde g} \approx 230$ GeV, allowing
better than $10\%$ tuning. UV sensitivity of the Higgs sector to a
heavy gluino is also reduced if the gluino gets its mass through
supersoft operators \cite{Fox:2002bu}. However, this requires adding
an extra $SU(3)$ adjoint fermion to the MSSM. Barring this
possibility, the most natural spectrum with a split third
generation has the gluino and sbottoms near $200-250$ GeV, stops
near $250-300$ GeV, and the first two generations of squarks near
$500$ GeV. Of course, no simple models exist that predict such a
spectrum.

\subsection{Tension with Gaugino Unification, Chargino Mass, and $b
  \rightarrow s \gamma$} The spectrum above, though natural, is badly
at odds with gaugino unification, which requires
$\frac{M_1}{\alpha_1}=\frac{M_2}{\alpha_2}=\frac{M_3}{\alpha_3}$.
Consequently, the optimistic bounds on the lightest chargino and
neutralino of $m_{\tilde \chi_1^+}\gtrsim 101$ GeV and $m_{\tilde
\chi_1^0}\gtrsim 46$ GeV respectively imply $m_{\tilde g} \gtrsim
450-500$ GeV for positive $\mu$, and $m_{\tilde g}\gtrsim 300-350$ GeV
for negative $\mu$ \cite{Chargino/Neutralino Bound}. With gaugino
unification, the $Z$ mass tuning must be at least $5-10\%$, regardless
of the Higgs sector.

An independent source of fine-tuning of $M_Z$ is the $\mu$-term.
Current limits are $\mu\gtrsim 140$ GeV or $\mu \lesssim -100$ GeV
from LEP chargino searches \cite{Chargino/Neutralino Bound}. The
numerical fine-tuning is negligible in either case, but again it is
striking that the charginos and neutralinos, which should be near
$M_Z$, are, like all other superpartners, \emph{above} it. In fact,
positive $\mu$ is preferred to match the observed $b \rightarrow s
\gamma $ branching ratio assuming minimal flavor violation, which is
maximally at odds with gaugino unification if the gluino is light
\cite{Flavor Violation}. In the NMSSM where the $\mu$-term is
generated dynamically, the seemingly mild requirement of
$|\mu|\gtrsim M_Z$ can be very difficult to explain without severe
tuning of parameters.

%% file: NmssmOverview.tex
\section{Hidden Tunings from a Hidden Higgs}\label{sec:nmssm}
In evaluating the naturalness of a SUSY model, all sources of tuning
should be accounted for. Models that address one source of
tuning may aggravate others. For this reason, it is important to
consider the new fine-tuning introduced by hiding or lifting the Higgs mass.

For concreteness, we restrict our attention to the NMSSM---the
simplest model in which the non-discovery of the Higgs at LEP is
consistent at tree-level.  The NMSSM has been advocated as a model
that has low fine-tuning in regions where the Higgs decays
predominantly to two lighter pseudoscalars or where the Higgs is
heavy at tree-level \cite{Dermisek:2005ar}.

The parameter space of the NMSSM is large and the sources of
fine-tuning differ from region to region, so a full analysis of the
fine-tuning is not practical. Instead we restrict our attention to two
regions where the NMSSM has the best naive chance of being natural.
These are: (i) large $\lambda$ so that the MSSM-like Higgs is heavy
and (ii) moderate $\lambda$ where a symmetry protects the mass of a
light pseudoscalar that can open up cascade decays, making a
significantly lighter Higgs consistent with LEP bounds.  The two
symmetries that accomplish this without decoupling the singlet are a
$U(1)_{PQ}$ and $U(1)_R$ \cite{R and PQ limits}. The tunings cited are
intended to clarify the importance of various sources of tension in
the models, and should be interpreted as order-of-magnitude estimates
only.

In each limit, we find that several phenomenological constraints
require tuning.  Obtaining viable electroweak symmetry breaking
constrains the model in both symmetry limits, though most severely in
$U(1)_{PQ}$. The $Z$-mass is once again somewhat fine-tuned, now
primarily because the soft-breaking parameter $A_\lambda$ must be
large to evade LEP chargino limits.  Moreover, unless $\lambda$ is
large, the singlino would have been seen at LEP. This limit has a
naturally light pseudo-Goldston boson, which is a candidate for
mediating cascade decays, but hiding the Higgs still requires mild
tuning of masses and mixing angles.  Electroweak symmetry breaking is
less tuned in the $U(1)_R$ limit.  This symmetry is, however, broken
by the Standard Model $A$-terms, so that the light pseudoscalar must
be tuned against radiative corrections.  We do not discuss dark matter or
flavor signals in detail, though they should be considered as well to
compare the NMSSM thoroughly to the MSSM.

We find that the Z-mass tuning in the NMSSM with a Higgs hidden by
cascade decays is mildly improved compared to the MSSM. As we will
show however, in the limits where cascade decays are natural,
obtaining viable electroweak symmetry breaking requires a tune in
the Higgs soft masses at the $\sim 1-10\%$ level. This problem is
generic in regions of parameter space that can hide the Higgs, and
elsewhere the tuning in the Higgs spectrum is at the $\sim 10\%$
level. Consequently, the NMSSM suffers from a fine-tuning problem of
its own.

Throughout this discussion, we assume that gaugino masses do not unify
and third-generation squarks are split from the first two generations,
allowing us to lower $m_{Q3}$ and $m_{U3}$ near $\approx 200$
GeV. Moreover, we will assume that the electroweak gaugino soft masses
$M_1$ and $M_2$ are decoupled except where otherwise noted.  This
loosens the constraints from chargino and neutralino searches at
LEP. These are poorly motivated assumptions, but they allow us to
divide the fine-tuning problem into manageable pieces.  We use the
Lagrangian conventions of \cite{Ellwanger:2004xm} throughout.

\subsection{Sources of Tuning in the NMSSM PQ Limit}
\subsubsection{Electroweak Symmetry Breaking}\label{sec:EWSB}
In this section, we attempt to resolve particular limits of the
NMSSM in which viable EWSB can occur radiatively--that is, for which
the origin is stable at $M_{GUT}$, but unstable at $M_Z$.

The potential for the Higgs and singlet vevs is given by
\begin{align}\label{NMSSMPotential}
    V(s,v_u,v_d)=&\lambda^2 (h_u^2 s^2+h_d^2 s^2+h_u^2 h_d^2)+\kappa^2
    s^4 -2 \lambda \kappa h_u h_d s^2 \nonumber \\
    & -2 \lambda A_\lambda h_u h_d
    s +\frac{2}{3}\kappa A_\kappa s^3 \nonumber + m_{H_u}^2 h_u^2 +
    m_{H_d}^2 h_d^2 \\
    &+ m_{S}^2 s^2 +\frac{1}{4} g^2 (h_u^2-h_d^2)^2,
\end{align}
where $g^2=\frac{1}{2} (g_1^2+g_2^2)$.  Assuming a given $\lambda$,
$\kappa$, $A_\lambda$, and $A_\kappa$, we wish to find the values of
soft masses $m_{H_u}^2$ and $m_S^2$ such that electroweak symmetry
breaking with $v_u^2+v_d^2=v^2 =(174 \mbox{GeV})^2$ is possible for
some suitable choice of $m_{H_d}^2$.  The size of this allowed
region relative to the radiative corrections to these soft masses
($\sim (200 \mbox{GeV})^2$) is a measure of the tuning for
electroweak symmetry breaking.  If $m_{H_u}^2$ or $m_{H_d}^2$ is
parametrically larger than $M_Z^2$, there is further tuning of the
weak scale.

The $U(1)_{PQ}$ under which $H_u, H_d$ and $S$ have charges 1, 1,
and -2 respectively requires $A_{\kappa}=0,\kappa=0$. In this case,
the extremization conditions on $V$ imply
\begin{eqnarray}\label{eq:PQExtremum}
  m_d^2 &=& \f{\mu A_\lambda}{\sin(2\beta)}-\lambda^2v^2-2\mu^2-m_u^2 ,\\
  M_Z^2 &=&
  \f{1-\cos(2\beta)}{\cos(2\beta)}m_u^2-\f{1+\cos(2\beta)}{\cos(2\beta)}m_d^2-2\mu^2,
  \\
  \mu   &=& \lambda s = \f{A_\lambda \sin (2 \beta)}{2(1+\xi)},
\end{eqnarray}
where $\xi=m_S^2/(\lambda^2 v^2)$, and corrections of order
$\f{\kappa}{\lambda}$ are expected. For small $m_S^2$ and nonzero
$h_u$ and $h_d$ vevs, the $\lambda A_{\lambda}$ term generates a
minimum at nonzero $s$. Thus, $A_\lambda$ controls both
$\mu_{eff}=\lambda s$ and $\mu B_{eff}=\lambda A_\lambda s$. Because
of this relation, $-M_Z^2/2 \lesssim m_{H_u}^2<0$ for small $m_S^2$.
$-(50 \mbox{GeV})^2 \lesssim m_S^2$ is also required for a stable
electroweak symmetry-breaking vacuum to exist with $\lambda\approx
0.7$.  $m_S^2 \gtrsim (30 \mbox{GeV})^2$ decreases $\mu$, and as we
discuss in Section \ref{inos}, raising $A_\lambda$ to compensate
results in fine-tuning of $M_Z$.  Because $m_{H_u}^2$ receives
radiative corrections of order $(100 \mbox{GeV})^2$ from stop soft
masses and gluinos, and $m_S^2$ receives corrections of comparable
size from $A_\lambda$ \cite{King:1995vk}, being in this region of
parameter space requires a tuning of a few percent.  The $m_{H_u}^2$
tuning has no analogue in the MSSM because there, $\mu$ and $\mu B$
can be varied independently.

If the universe was ever at a temperature of $T\sim 100$ GeV, the
negative $m_S^2$ is cosmologically dangerous.  Although the
electroweak symmetry breaking minimum is the global minimum of the
potential, it is not necessarily the minimum to which a hot early
universe rolls first. At a temperature $T\gtrsim M_Z$, the leading
finite-temperature corrections to the soft masses are given by
\cite{Weinberg:1974hy}
\begin{eqnarray}
m_{H_u}^2(T) &\approx& m_{H_u}^2 + \frac{T^2}{12} \f{6 y_t^2}{\sin(\beta)^2}\\
m_{S}^2(T) &\approx& m_{S}^2 + \frac{T^2}{12} 4 \lambda^2 .
\end{eqnarray}
Consequently, we expect the singlet field to roll into the local
extremum along the $s$-axis (which is typically a local minimum, not
a saddle point) before the Higgs field rolls off whenever $m_S^2
\lesssim \frac{4\sin(\beta)^2 \lambda^2}{6 y_t^2} m_{H_u}^2$. The
timescale for tunneling to the true EWSB vacuum is typically much
longer than Hubble time scale $H(T)^{-1}$ at $T\sim M_Z$. If the
cosmological constant is tuned to cancel the vacuum energy in the
electroweak minimum, then the slow decay rate allows the universe to
enter an inflationary phase below the weak scale. A detailed study
of this cosmology would be necessary to check the viability of this
scenario, but we assume that it is generically unacceptable.  This
argument implies further fine-tuning of $m_S^2$ for cosmologically
safe EWSB.

\begin{figure}[htbp]
\includegraphics[width=3in]{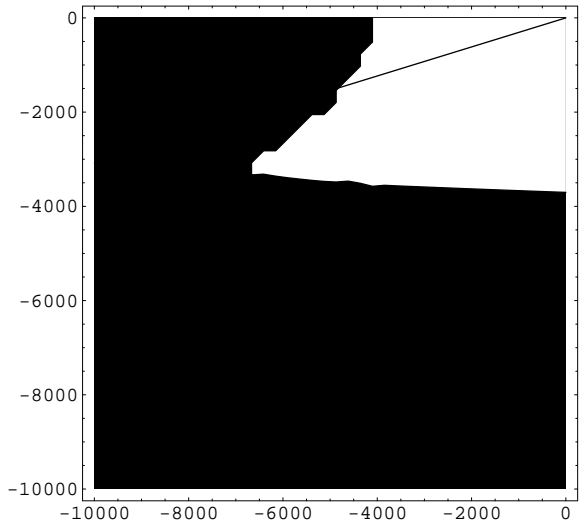}
\caption{White region: Allowed soft mass ranges for electroweak
symmetry
  breaking with $M_Z=91$ GeV for some $m_{H_d}^2$, in the PQ limit
  of the NMSSM, with parameters $\lambda=0.68$, $\kappa= 0.03$,
$A_\lambda= 350$, and $A_\kappa=50$.  Below the black line, an
electroweak symmetry-breaking global minimume exists, but
cosmologically viable electroweak symmetry breaking is not
guaranteed.}\label{fig:ewsb-PQ}
\end{figure}

Figure \ref{fig:ewsb-PQ} shows the constraints on the Higgs soft
masses for electroweak symmetry breaking in a typical slice of the
PQ-symmetric limit.  The white region indicates ranges of
$m_{H_u}^2$ and $m_S^2$ for which an electroweak symmetry-breaking
vacuum exists, for some $m_{H_d}^2$, with $M_Z=91$ GeV and $\tan
\beta>1$.  In the black regions, no stable symmetry-breaking minimum
exists that can reproduce $M_Z$.  The calculations are done at
tree-level.  The line through the plot indicates the boundary of
cosmological safety.

\subsubsection{Neutralino and Chargino Masses\label{inos}}
The NMSSM introduces a new neutralino and ties the mass of the
chargino to the couplings of the Higgs sector in new ways.  It is
important, therefore, to consider the implications of experimental
bounds on charginos and neutralinos for allowed regions of NMSSM
parameter space.  Charginos are constrained by the direct production
bounds from LEP II \cite{Chargino/Neutralino Bound,Eidelman:2004wy},
$m_{\chi^\pm} \gtrsim 95-105$ GeV. Neutralinos must evade bounds
from the $Z$-pole width and associated production of two neutralinos
off-shell.  They must also exceed roughly $m_H/2$ so that Higgs
invisible decays are not allowed.  We consider each of these
constraints in the $PQ$ limit, and all expressions are corrected by
terms of higher order in $\kappa$.

As discussed in the previous section, the LEP chargino bounds
require $\mu \lesssim -100$ GeV or $\mu \gtrsim 140$ GeV. In the
approximate-$PQ$, small-$m_S$ regime,
\begin{equation}\label{muterm}
\mu \approx \frac{A_\lambda \sin (2 \beta)}{(1+\xi)}
(1+\mathcal{O}(\kappa)) .
\end{equation}
From eqs. (\ref{eq:PQExtremum}) and (\ref{muterm}), the effective
$\mu B$-term is $\mu B_{eff}=\mu A_{\lambda}$ and the $\mu$-term is
controlled by $A_{\lambda}$. As $A_{\lambda}$ is increased to make
$\mu$ sufficiently large, the first equation in
(\ref{eq:PQExtremum}) shows that either $m_d^2$ or $m_u^2$ must also
increase. If electroweak symmetry breaking occurs radiatively, then
$m_u^2\sim M_Z^2$. Assuming this condition, large $A_{\lambda}$
requires a compensating large $m_d^2$ and increasing $\tan \beta$ .
Thus, large $A_{\lambda}\gtrsim 250-350$ GeV required to make
$|\mu|\gtrsim 100-140$ GeV introduces a quadratic tuning of order
$\sim 10\%$ to keep the $Z$ light.

The lightest neutralino bounds place a further constraint on NMSSM
parameters. Ignoring mixing with gauginos, the lightest higgsino
mass is given by
\begin{equation}
m_{\chi_0^1}\approx \frac{\lambda^2 (2 \gamma \mu^2+ v^2 \sin
  2\beta)}{\sqrt{\lambda^2 v^2+\mu^2}} + O\left(\left(\frac{v}{s}\right)^2
  \gamma^2+\left(\frac{v}{s}\right)^4\right),
\end{equation}
where the small parameter $\kappa/\lambda \equiv \gamma$ quantifies breaking of the $PQ$ symmetry.
With $\gamma << 1$, the chargino and neutralino constraints together imply an approximate lower bound on
$\lambda$:
\begin{equation}
\lambda \gtrsim \sqrt{\frac{m_{\chi_0^1,min}\mu} {v^2}} \approx 0.5
,\label{eqn:approxSinglino}
\end{equation}
for $m_{\chi_0^1,min}\approx 45$ GeV, the rough bound from direct
$Z$-pole production. Given that large $A_{\lambda}$ implies small
$\sin(2\beta)$, larger $\lambda$ is required to keep
$m_{\chi_0^1}>\f{M_Z}{2}$ GeV in most of the parameter space of the
$PQ$ limit. It should also ne noted that the bound from
$h->\mbox{invisible}$ typically requires heavier $\chi_0^1$.

Before considering the requirements for cascade decays, let us
summarize the constraints on the PQ limit imposed by non-Higgs
phenomenology: Large $A_\lambda$ is required to evade LEP chargino
bounds, and in turn imposes a tuning to get a light Z of about $\sim
10\%$. This is comparable to the tuning imposed by gluino
constraints. Considering the neutralino mass bound as well forces us
into the regime $\lambda\gtrsim 0.5$. Such large $\lambda$ raises
the mass of the Higgs to $\approx 105$ GeV. Thus, in the PQ limit,
where cascade decays could naturally hide the MSSM-like Higgs, the
Higgs is always fairly heavy when neutralino and chargino
constraints are satisfied. The only way to avoid this is to back out
of the PQ limit, which reintroduces problems with EWSB and requires
additional tuning to keep the pseudoscalar light.

\subsubsection{Higgs Sector: Cascade Scenario}
The experimental constraints on the MSSM with singlets added have
recently been discussed in \cite{Chang:2005ht}. The most
constraining bounds are on Higgs-strahlung production with decay
modes $h \rightarrow b \bar b$ and $h \rightarrow \mbox{invisible}$.
For SM-like Higgs-strahlung cross sections, these bounds exclude
$m_h<114.4$ GeV for $h \rightarrow b \bar b$ and $m_h<114$ GeV for
$h \rightarrow \mbox{invisible}$. This assumes $100\%$ branching to
the decay mode in question. To hide a Higgs below $114$ GeV, either
the cross section for Higgs-strahlung and direct production must be
lowered, or a visible but less constrained cascade decay channel of
the form $h \rightarrow \phi \phi \rightarrow 4 X$ must be opened
up. Of course, this requires a lighter boson $\phi$ (scalar or
pseudoscalar) with $m_\phi\lesssim m_h/2 \lesssim 57$ GeV.

A light scalar $s$ is further constrained by direct bounds on
Higgs-strahlung $Z \rightarrow s Z$ production, while a light
pseudoscalar $a$ is constrained by bounds on direct production $Z
\rightarrow h a$. The suggestive notation $s$ and $a$ refers to the
lightest scalar or pseudoscalar of the Higgs spectrum, which must be
primarily singlet to satisfy $m_\phi< m_h/2$ and be allowed by LEP,
but will be mixed.

The cascade decays $h \rightarrow \phi \phi \rightarrow 4 b$ are
further constrained for $m_h\lesssim 110$
\cite{Chang:2005ht,Ellwanger:2004xm}. Hence a hidden Higgs requires
either suppression of Higgs-strahlung (by a factor of $\sim 10$ over
SM levels) or $m_\phi\lesssim 11$ GeV, so that $4b$ production is
kinematically suppressed, and the less constrained $h \rightarrow
\phi \phi \rightarrow 4 \tau$ decay mode dominates.

In either the $4b$ or the $4\tau$ scenario, the Higgs sector appears
tuned to keep a pseudoscalar light. This tuning is typically at the
$1-10\%$ level to keep $m_\phi\lesssim 11$ GeV. If, however, the
intermediate Higgs $\phi$ of the cascade is protected by an
approximate symmetry, this tuning can be reduced. The $PQ$
pseudo-Goldstone boson is a good candidate, as the $PQ$ limit is
also favored for EWSB.

The mass of the lightest pseudoscalar in the $PQ$ limit is given by
\begin{equation}
m_{A_1}^2 \approx \gamma \frac{6 \mu^2 (3 \sin 2 \beta
  \lambda^2 v^2 - 2 A_\kappa \mu)}{4 \mu^2+ \lambda^2 v^2 \sin^2
  2\beta} + \mathcal{O}(\gamma^2 \mu^2)
\end{equation}
We note that the pseudoscalar can be made light by tuning $A_\kappa$,
 even when $\kappa$ is only moderately small. From the
 lower bounds $\lambda\gtrsim 0.5$ and $\mu \gtrsim 140$, the natural
 size of $m_{A_1}$ is
\begin{equation}
m_{A_1}^2 \gtrsim \frac{3\gamma}{2} (3 \sqrt{\mu_{min}^2+ \lambda^2
  v^2}\, m_\chi^{min} \lambda^2),
\end{equation}
or $m_{A_1} \sim \sqrt{\lambda \kappa}\, (250 \mbox{GeV})$.
When $\f{\lambda}{\kappa} < .04$, $m_A \lesssim 50$ GeV is natural
and the cascade $h \rightarrow 2  \rightarrow 4b$ opens up.  When
$\f{\lambda}{\kappa} \lesssim .002$, $m_A< 11$ GeV and $h
\rightarrow 2 a \rightarrow 4 b$ is kinematically forbidden, and the
pseudoscalar naturally decays to $2 \tau$.

The necessity of large $\lambda$ in the hidden Higgs scenarios
discussed below suggest that, in the $PQ$ limit, the heavy Higgs is
most natural.  In fact, a heavy Higgs can be hidden quite
generically for $\lambda \gtrsim 0.65$ in the $PQ$ limit. The
tension with $A_\lambda$ and EWSB remains in this case, however.

\subsubsection{Benchmark Parameter Scans}
The plots below are generated using NMHDECAY 1.1 to study the
phenomenology of the NMSSM for various choices of parameters
\cite{Ellwanger:2004xm}. Inputs are pre-processed by a Mathematica
script that finds the derived parameters $\mu$ and $\tan \beta$,
which are inputs to NMHDECAY, for given $m_{H_u}^2$ and $m_S^2$.  As
above, $m_{H_d}^2$ is determined by fixing $M_Z$.  The Mathematica
front-end determines $\mu$ and $\tan \beta$ at tree-level, and
NMHDECAY calculates at one-loop order, so the input values of
$m_{H_u}^2$ and $m_S^2$ are not the same as those returned by
NMHDECAY. In the tables below, we cite the input values.  The ranges
of $m_{H_u}^2$ ($m_S^2$) returned by NMHDECAY are roughly 500 (40)
GeV$^2$ larger.

\begin{figure}[htbp]
\includegraphics[width=3in]{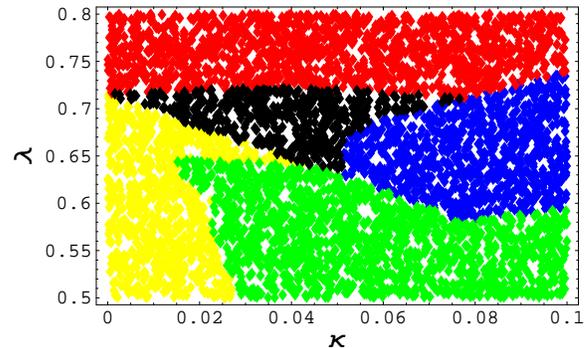}
\caption{Phenomenology constraints on the NMSSM as $\lambda$ and
  $\kappa$ are varied.  Fixed parameters are listed in Table
  \ref{tab:phenonumbers}.  The red region at the top and black region
  just below it are phenomenologically allowed, with the
  second-heaviest (MSSM-like) Higgs heavier and lighter than 114.4 GeV,
  respectively.  The center-right blue region is excluded
  by Higgs decays to nonsupersymmetric particles.  The yellow region
  on the lower left is excluded by light neutralino
  searches, while the lower-right green region is excluded by both
  neutralino searches and Higgs decay searches.}\label{fig:marg}
\end{figure}

\begin{figure}[htbp]
\includegraphics[width=3in]{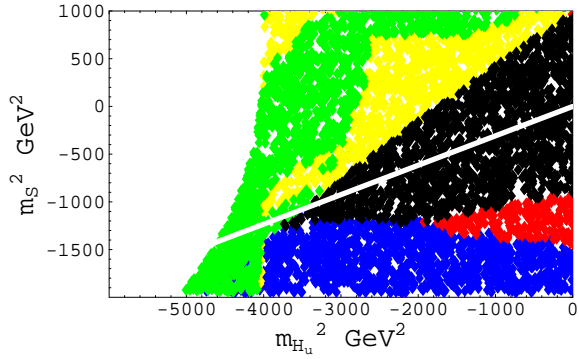}
\caption{Phenomenology of regions of $m_{H_u}^2$-$m_S^2$ parameter
  space where EWSB consistent with $M_Z$ is possible.  Fixed
  parameters are listed in Table \ref{tab:phenonumbers}.  The black
  region on the right is phenomenologically allowed, and has the
  second-heaviest (MSSM-like) Higgs hidden below 114.4 GeV.  The small
  red region directly below it is also allowed, but with the MSSM-like
  Higgs mass $\ge$ 114.4 GeV.  The lower blue region is excluded by
  Higgs decays to nonsupersymmetric particles.  The yellow regions on
  the upper left and middle are excluded by light neutralino searches,
  while the $\Gamma$-shaped green region is excluded by both
  neutralino searches and Higgs decay searches.  Only points above the
  white line are cosmologically viable if symmetry is restored at high
  temperatures.  Points below the line roll first along the singlet
  axis, and typically do not roll off into an electroweak
  symmetry-breaking vacuum }\label{fig:soft}
\end{figure}

In Figures \ref{fig:marg} and \ref{fig:soft}, we have scanned two
parameters randomly while leaving the other four fixed, to
demonstrate the dominant sources of fine-tuning in the
$PQ$-symmetric limit.  The parameter values and ranges used to
generate these figures are given in Table \ref{tab:phenonumbers}.
Scanning the soft masses reveals the same contour as in Figure
\ref{fig:ewsb-PQ} for electroweak symmetry breaking.  A substantial
region near $m_S^2=0$ is consistent with phenomenology.  Varying
$\lambda$ and $\kappa$ reveals that large $\lambda$ is necessary to
evade neutralino constraints, and that a hidden Higgs is not generic
except in the PQ limit.  For $\lambda \gtrsim 0.7$, the MSSM-like
Higgs is heavier than 114.4 GeV, and its decays are unconstrained.
But this region, too, becomes smaller as $\kappa$ increases beyond
$\approx 0.1$, because the Higgs searches constrain the lightest
Higgs as well away from the PQ limit.

\begin{table}
\begin{tabular}{|c|c|c|c|c|c|c|c|c|}
\hline Figure & $\lambda$ & $\kappa$ & $A_\lambda$ & $A_\kappa$ &
$m_{H_u}^2$ & $m_S^2$\\
\hline
\ref{fig:marg} & 0.5-0.8 & 0.0-0.1 & 350. & 50. & -1500. & 0  \\
\ref{fig:soft} & 0.68 & 0.03 & 350. & 50. & * - 0. & -2000 -- 1000 \\
\hline
\end{tabular}
\caption{Fixed parameter values and scanned ranges for Figures
  \ref{fig:marg} and \ref{fig:soft}.  All dimensionful quantities are
  given in $GeV$, and in the second line, * denotes that the lower
  limit of the scanned range is not input by hand but dictated by
  electroweak symmetry breaking.}\label{tab:phenonumbers}
\end{table}


\subsection{Sources of Fine-Tuning in the NMSSM R Limit}
\subsubsection{Electroweak Symmetry Breaking}
The $U(1)_{R}$ limit where $H_u,H_d$ have charge 1 and $S$ is
neutral corresponds to $A_{\lambda}=A_{\kappa}=0$.  With $h_u^2+h_d^2=v^2$,
the singlet potential is stable at the origin unless $m_S^2\lesssim
-(\lambda^2v^2-\lambda\kappa v^2 \sin(2\beta))$. When the singlet
origin is destabilized,
\begin{eqnarray}
  s^2 &=& -\f{(m_S^2+\lambda^2v^2-\lambda\kappa v^2 \sin(2\beta))}{2\kappa^2} ,\\
  \mu &=& \lambda s =
  \f{-\lambda}{\kappa}\f{(m_S^2+\lambda^2v^2-\lambda\kappa v^2 \sin(2\beta))}{\sqrt{2}} .
\end{eqnarray}
Thus, very large negative $m_S^2$ is necessary to generate a
sufficiently large $\mu$-term. In this case, the singlet vev rolls
away from the origin first, but $h_u$ and $h_d$ can be unstable
along the $s$-axis.  The Higgs directions are unstable at the
$s$-minimum at $s^2=-\frac{\lambda^2}{2 \kappa^2} m_S^2$ for
$m_{H_u}^2+m_{H_d}^2 \lesssim 0$ (this expression is valid for
$\lambda \approx \kappa$, $|m_{H_u}^2|,|m_{H_d}^2|\ll |m_S^2|$.
This, too, is consistent with radiative symmetry breaking.  Although
$m_S^2$ must be quite large and negative at high scales, the vacuum
on the singlet axis is stable until $m_{H_u}^2+m_{H_d}^2$ runs
negative.

\begin{figure}[hbtp]
\includegraphics[width=3in]{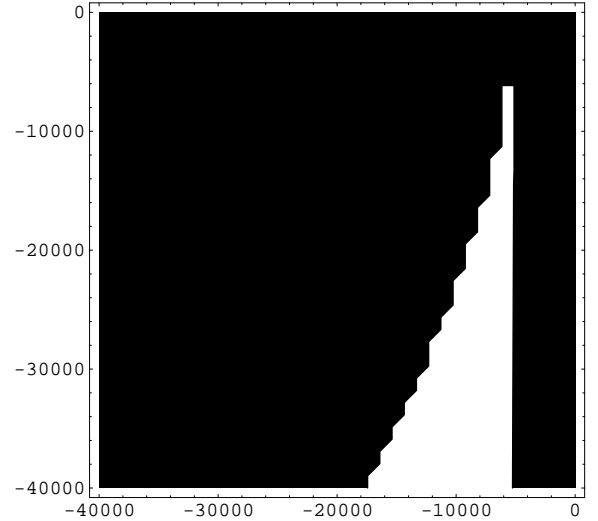} \caption{White region: Allowed
soft mass ranges for electroweak symmetry
  breaking with $M_Z=91$ GeV for some $m_{H_d}^2$, in the
R-symmetric limit
  of the NMSSM, with $\lambda=\kappa=0.63$, $A_\lambda=0.5$, $A_\kappa=0$.}\label{fig:ewsbR}
\end{figure}

Figure \ref{fig:ewsbR} shows the regions of Higgs soft masses for
which electroweak symmetry breaks in a typical slice of the $U(1)_R$
limit of the NMSSM.  The white regions indicate values of
$m_{H_u}^2$ and $m_S^2$ for which a tree-level electroweak
symmetry-breaking vacuum exists, for some $m_{H_d}^2$, with $M_Z=91$
GeV and $\tan \beta>1$.  In the black regions, no stable
symmetry-breaking minimum exists that can reproduce $M_Z$.
$m_{H_u}^2$ must be somewhat smaller than $\sim (200 \mbox{GeV})^2$,
the size of its radiative corrections, so roughly 10\% tuning is
required for electroweak symmetry breaking.

\subsubsection{Higgs Mass Constraints}
The chargino and neutralino tensions that are problematic in the PQ
limit are unimportant in the R-symmetric limit, where the higgsinos
are generally heavier.  The primary source of tuning in the
R-symmetric limit is Higgs decays.  In particular, whereas it is
technically natural to assume $m_{A_1}$ small in the PQ-symmetric
limit because of the approximate symmetry, $U(1)_R$ is broken by the
Standard Model $A$-terms and gaugino masses, so that $A_\lambda$
receives radiative corrections of order $100-200$ GeV.  To evade LEP
bounds without significantly tuning the Higgs couplings to the Z
small, a Higgs below 112 GeV must decay to 4$\tau$, requiring
$m_{A_1} < 12$ GeV.

The mass of $U(1)_R$ pseudo-scalar is
\begin{equation}\label{Rscalar}
    m_{A_1}^2\approx \f{9\lambda^2v^2A_{\lambda}}{2\mu}-\f{3\kappa\mu A_{\kappa}}{\lambda}
\end{equation}
so that tuning of $A_{\lambda}$ against $A_{\kappa}$ is necessary.
Given that radiative corrections to $A_{\lambda}$ are $\delta
A_{\lambda}\sim 100$ GeV when the Higgs sector soft masses are large
and the cancellation required is $\sim 0.5$ GeV, this tuning is at
the $\sim 1-10\%$ level. Cascade decays to 4b are severely
constrained by the data \cite{LEPmssmHiggs}, so tuning to suppress
Higgs-strahlung of order $\sim 1-10\%$ or more is required.

In \cite{Other NMSSM hidden higgs references}, it was claimed that
for small $\lambda\sim\kappa\sim 0.2$, $|A_{\lambda}|\sim 100$ GeV,
$A_{\kappa}\sim 3$ GeV, and moderate $\tan(\beta)$, there are
technically natural points in the R-limit that hide a MSSM-like
Higgs near $\approx 100$ GeV. The only fine-tuning required there is
in obtaining electroweak symmetry breaking. If the LEP excess of
$2b$ events near this mass range is a signal, then the parameter
choices required to explain the spectrum in these regions are
constrained only by the positive signal. The tunings we consider are
necessary to \emph{avoid} predicting a signal, and therefore should
be considered phenomenologically unnatural.

For a heavier Higgs (which requires $\lambda \gtrsim 0.63$),
cascades to 4b are not constrained, and this essentially reduces to
the large quartic scenarios of \cite{Adding Quartics}. In these
cases, the remaining tuning is dominated by requiring viable
electroweak symmetry breaking ($\sim 5-10\%$).

\begin{figure}[htbp]
\includegraphics[width=3in]{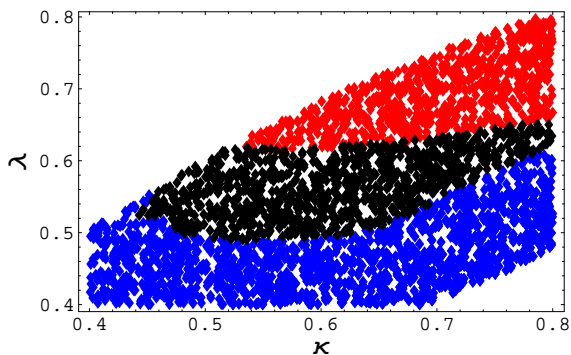}
\caption{Phenomenology constraints on the NMSSM as $\lambda$ and
  $\kappa$ are varied in the R-symmetric limit.  Fixed parameters are
  listed in Table \ref{tab:R-params}.  The red region at the top
  and black region just below it are phenomenologically allowed, with
  the lightest Higgs heavier and lighter than 112
  GeV, respectively.  The blue region at the bottom is excluded by
  Higgs decays.}\label{fig:R1}
\end{figure}

\begin{figure}[htbp]
\includegraphics[width=3in]{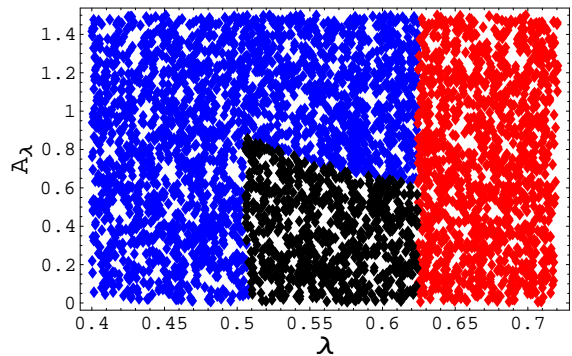}
\caption{Phenomenology constraints on the NMSSM as $\lambda$ and
  $A_\lambda$ are varied in the R-symmetric limit.  Fixed parameters
  are listed in Table \ref{tab:R-params}.  The red region at the right
  and black region in the center-bottom are phenomenologically
  allowed, with the lightest Higgs heavier and lighter than 112 GeV,
  respectively, and the blue region at the left is excluded by
  Higgs decay searches.}\label{fig:R2}
\end{figure}

Figures \ref{fig:R1} and \ref{fig:R2} display representative
cross-sections of the parameter space, with the allowed parameter
space for a heavy ($> 112$ GeV, red) or light hidden Higgs ($< 112$
GeV, black), and the region where the $Higgs$ is excluded by Higgs
decay searches at LEP (blue).  The parameter ranges used are given
in Table \ref{tab:R-params}.  All points allowed by the experimental
constraints have Landau poles below $M_{GUT}$.

\begin{table}
\begin{tabular}{|c|c|c|c|c|c|c|c|c|}
\hline Figure & $\lambda$ & $\kappa$ & $A_\lambda$ & $A_\kappa$ &
$m_{H_u}^2$ & $m_S^2$\\
\hline
\ref{fig:R1} & 0.4-0.8 & 0.4-0.8 & 0.5 & 0.0 & $-100^2$ & $-200^2$  \\
\ref{fig:R2} & 0.4-0.8 & 0.65 & 0.0-1.5 & 0.0 & $-100^2$ & $-200^2$ \\
\hline
\end{tabular}
\caption{Fixed parameter values and scanned ranges for Figures
  \ref{fig:R1} and \ref{fig:R2}.  All dimensionful quantities are
  given in $GeV$.}\label{tab:R-params}
\end{table}

%% file: conclusion.tex
\section{Conclusions}\label{sec:conclusion}
In this paper, we have reviewed the constraints on the naturalness
of low-scale SUSY imposed by experimental limits.  Many recent
attempts to address the little hierarchy problem have focused on
modifying the MSSM Higgs sector, but it is unclear whether such
modifications reduce the fine-tuning required for a consistent SUSY
completion of the Standard Model or merely move the fine-tuning
around.

Experimental bounds on the squark, gluino, and electroweak gaugino
sectors all suggest fine-tuning of $M_Z$.  This tuning exists
independently of the Higgs sector, but can be alleviated by
abandoning gaugino unification and splitting the third-generation
squark masses from the first and second generations.

In addition to the fine-tuning independent of Higgs physics, we
considered tuning associated with a modified Higgs sector, taking
the NMSSM as an example and further specializing to the $PQ$- and
R-symmetric limits with positive $\mu$ to avoid
$\mbox{Br}(b\rightarrow s\gamma)$ constraints \cite{Flavor
Violation}. These limits can naturally open up cascade decay
channels for the MSSM-like Higgs, permitting a hidden Higgs.

In the PQ-symmetric limit, tuning is required to evade Higgs bounds,
though not directly worse than in the MSSM. Cosmologically safe
radiative electroweak symmetry breaking requires that the soft
masses be tuned small at the $\sim 1-10\%$ level. The $Z$-mass
depends on $A_\lambda$, which must be large to raise the mass of
charged higgsinos above experimental limits.  This generates an
independent tuning of $\sim 10\%$. Further tension comes from
evading direct searches for the Higgs and the lightest neutralino,
which in this case is mostly singlino.  The neutralino bounds are
only satisfied for relatively large $\lambda\gtrsim 0.5$, so that
the MSSM-like Higgs state must be heavier than $100$ GeV. With small
$\kappa$, this generically keeps all Landau poles above the
unification scale.

In the R-symmetric limit, large $\kappa,\lambda\gtrsim 0.6$ is
typically required so a Landau pole below the unification scale is
typically present. To generate a sufficiently large $\mu$ term, the
singlet soft mass must be large and negative, causing mild tension
with electroweak symmetry breaking. Consequently, tuning in Higgs
sector soft masses at the $\sim 10\%$ level is required for
electroweak symmetry to break. In cases where the lightest MSSM-like
Higgs states are below $\approx 110$ GeV, the A-terms are required
to be unnaturally small or tuned against each other in order to
evade Higgs search constraints. This introduces a tuning at the
$\sim 1-10\%$ for parameter ranges that would generically go unseen.
This is not apply to the points discussed in \cite{Other NMSSM
hidden higgs references} that are used to explain the excess of $2b$
events associated with a Higgs mass of $\approx 100$ GeV.

Away from the PQ- or R-symmetric limit, other regions of NMSSM
parameter space may allow more natural $M_Z$, but require comparable
or worse tuning for light pseudoscalars or electroweak symmetry
breaking.

As the NMSSM illustrates, the lack of technical naturalness required
by demanding a Z at the bottom of the SUSY spectrum can be partially
exchanged for tunings in the Higgs and higgsino spectra. It is not
clear whether this fine-tuning is less mysterious than $Z$-mass tuning
or whether it is a comparable cause for concern.

Tuning peculiar to the NMSSM may be reducible by further modifying
the Higgs sector, though certain extensions may merely obscure this
tuning, so they must be considered with care.  For example, an
explicit $\mu$ term allows smaller $A_\lambda$ in the PQ-limit and
so can reduce the $Z$-mass tuning, but may also increase the tuning
to open cascade decay channels.  Other MSSM+singlet models suggested
in \cite{Chang:2005ht} may reduce tuning more successfully, but they
too must be considered in detail to address this question.

The possibility of a tuned but light superpartner spectrum also
merits study because it is a very different scenario for LHC
physics.  The light spectrum is clearly promising, as even the
weakly interacting superpartners will easily be within reach of the
LHC.

By itself, the NMSSM does not eliminate the fine-tuning of SUSY, but
it does change the nature of this tuning and suggest an alternate
direction for model-building. If the conventional scenario of heavy
superpartners is correct, solutions to the little hierarchy problem
must explain the relative lightness of the $Z$ and Higgs. If the
spectrum is light, models will still need to explain the surprising
order of the spectrum and the fine-tuning of the $Z$-mass suggested by
current superpartner bounds. They must also, of course, consistently
explain electroweak symmetry breaking, heavy higgsinos, and Higgs
phenomenology that is invisible to LEP.  Contrary to some claims, the
NMSSM does not accomplish these tasks without a fine-tuning problem of
its own. It would be interesting to construct models that do hide the
Higgs states from detection at LEP and do not suffer from this
problem.